# Electric field measurements of Rydberg atomic frequency comb based on pulsed laser excitation

**KE DI,[1,2,3] CHENGLIN YE,[1,2] YIJIE DU[4], YU LIU[1,2], FENG GAO[5], JIAJIA DU[1,2] AND JUN HE[3*]**

[1]*Chongqing University of Post and Telecommunications, Chongqing 400065, China*
[2]*Chongqing Engineering Research Center of Intelligent Sensing Technology and Microsystem, Chongqing University of Post and Telecommunications, Chongqing 400065, China*
[3]*State Key Laboratory of Quantum Optics Technologies and Devices, Shanxi University, Taiyuan 030006, China*
[4]*State Key Laboratory of Space Information System and Integrated Application, and Beijing Institute of Satellite Information Engineering, Beijing, China*
[5]*Key Laboratory of Time Reference and Applications, National Time Service Center, Chinese Academy of Sciences, Xi'an 710600, China*
**author* hejun@sxu.edu.cn

**We present an innovative frequency comb methodology utilizing pulsed lasers for Rydberg atoms and implement it for electric field measurement. It achieves the Rydberg state population of multi-velocity group atoms through the two-photon resonant excitation of a 509 nm pulsed laser and an 852 nm continuous laser. The frequency comb approach markedly elevates the population of Rydberg atoms and augments the atomic density for sensing, thereby enhancing measurement sensitivity. Our investigations generated high-sensitivity measurements of electric fields across a broad spectrum from 20 kHz to 96 MHz, with a minimum measured electric field sensitivity of 2.9 $\mu V/cm/Hz^{1/2}$. Additionally, we have exhibited a high degree of measurement sensitivity in the 66 MHz and 88 MHz broadcast communication frequencies. This research enhances the effective detection of microwave signals over a broad spectrum of frequency bands utilizing Rydberg atoms and introduces an innovative technical methodology for microwave metrology grounded in Rydberg atoms.**

To Rydberg atoms have seen breakthroughs in precision measurements, attributed to their large orbital radii and high polarizabilities [1-4]. These characteristics confer notable quantum coherence [5], enhancing atomic clock accuracy [6] and enabling miniaturized magnetometers [7]. In broadcast communications, the high electric field measurement sensitivity of Rydberg atoms is vital for detecting weak signals [8]. This tech advance boosts signal detection ability and improves communication system stability and reliability, profoundly influencing communication technology development [9-12].

In recent years, pulsed light technology has advanced significantly in Rydberg atom excitation. Research teams have used pulsed lasers to achieve time domain Ramsey interference measurements of electron coherence's ultrafast evolution [13], observe ionization and ultracold plasmas [14], study ultrafast many body dynamics [15], and excite rubidium atoms to Rydberg states [16].Based on the room temperature cesium atom system [17], researchers first developed a Rydberg atom microwave frequency comb spectrometer, opening a new way for quantum precision measurement in the microwave band [18].However, the achieved frequency comb isn't a true Rydberg atom frequency comb as its generation mainly depends on external microwave field modulation. This method needs to control multiple microwave fields (MFCs) simultaneously and relies on complex laser frequency stabilization and differential detection, increasing system cost and operation difficulty. Also, its maximum measurement range is only 125 MHz, unable to cover higher frequency ranges like terahertz, thus limiting applications in quantum measurement and high precision spectroscopy.

In this paper, we introduce a novel Rydberg atom-based frequency comb technique that capitalizes on the frequency selectivity advantage of pulsed light technology. Utilizing the Doppler effect, atoms with varying velocities may be stimulated to the Rydberg state by two-photon resonance. The Rydberg state preparation of multi velocity group atoms can be facilitated by the selective excitation of specific velocity group atoms through the adjustment of the pulsed laser repetition frequency. This substantially increases the quantity of Rydberg state atoms and establishes an authentic Rydberg atom frequency comb.

We have performed theoretical analysis and experimental validation to achieve an authentic Rydberg atom frequency comb under pulsed light excitation and conducted electric field measurements. In the experiment, we established a minimum electric field measurement sensitivity of roughly 2.9$\mu V/cm/Hz^{1/2}$, and attained a measurement sensitivity of microvolts per centimeter at the critical broadcast communication frequencies of 66 MHz and 88 MHz, with a precision of around 1.27 Hz. The generation of Rydberg atoms using pulsed laser excitation offers distinct advantages, positioning this technique as a crucial platform for enhancing measurement sensitivity in precision measurement.

Fig. 1(a) illustrates the energy level diagram of the Rydberg state of cesium ($^{133}Cs$), including the ground state $|g\rangle$, the intermediate state $|i\rangle$, and the Rydberg state $|r\rangle$. Two laser beams resonantly link the ground state $|g\rangle$ and the Rydberg state $|r\rangle$. Let $\omega_p$ and $\omega_c$ represent the angular frequencies of the probing light and the coupling light, respectively. $\Delta_p$ and $\Delta_c$ denote the frequency detuning of the probing light and the coupling light in relation to the atomic resonant transition. The associated Rabi frequencies are $\Omega_p = \mu_{21}E_p/(2h)$ and $\Omega_c = \mu_{32}E_c/(2h)$ , where $\mu_{21}$ and $\mu_{32}$ denote the electric dipole moments of the respective transitions, and $\gamma_{21}$ and $\gamma_{32}$ represent the decay rates of the 6P excited state and the Rydberg state R.

The Hamiltonian of the interaction between the atom and the light field is:

$$H = H_a + H_l + H_{al} \tag{1}$$

Within this context, $H_a$ denotes the atomic Hamiltonian, $H_l$ represents the light field Hamiltonian, and $H_{al}$ signifies the interaction Hamiltonian.

The Hamiltonian of the three-level atomic system could be written in matrix form using the rotating wave and dipole approximations as follows:

$$H = \frac{\hbar}{2}\begin{pmatrix} 0 & \Omega_p & 0 \\ \Omega_p & -2\Delta_p & \Omega_c \\ 0 & \Omega_c & -2(\Delta_p + \Delta_c) \end{pmatrix} \tag{2}$$

The temporal development of the density matrix characterizes the progression of the three-level system, with its dynamics articulated by the Lindblad master equation:

$$\dot{\boldsymbol{\rho}} = -\frac{i}{\hbar}[H, \boldsymbol{\rho}] + \boldsymbol{L}(\boldsymbol{\rho}) \tag{3}$$

Within this context, $\boldsymbol{L}(\boldsymbol{\rho})$ denotes the dephasing matrix, whereas $\boldsymbol{\rho}$ represents the density matrix of the atom:

$$\boldsymbol{\rho} = \begin{pmatrix} \rho_{11} & \rho_{12} & \rho_{13} \\ \rho_{21} & \rho_{22} & \rho_{23} \\ \rho_{31} & \rho_{32} & \rho_{33} \end{pmatrix} \tag{4}$$

In the three-level atomic system, the dephasing matrix $\boldsymbol{L}(\boldsymbol{\rho})$ for spontaneous emission is expressed as:

$$\boldsymbol{L}(\boldsymbol{\rho}) = \begin{pmatrix} \gamma_{21}\rho_{22} & -\frac{\gamma_{21}}{2}\rho_{12} & -\frac{\gamma_{32}}{2}\rho_{13} \\ -\frac{\gamma_{21}}{2}\rho_{21} & \gamma_{32}\rho_{33} - \gamma_{21}\rho_{22} & -\frac{\gamma_{21}+\gamma_{32}}{2}\rho_{23} \\ -\frac{\gamma_{32}}{2}\rho_{31} & -\frac{\gamma_{21}+\gamma_{32}}{2}\rho_{32} & -\gamma_{32}\rho_{33} \end{pmatrix} \tag{5}$$

In the equation, $\gamma_{32}$ represents the decay rate from the Rydberg state to the intermediate state, whereas $\gamma_{21}$ denotes the decay rate from the intermediate state to the ground state.

Fig. 1(b) illustrates the contributions of atoms within various velocity groups to the EIT transmission signal. In the theoretical simulation, the intensity of the probe field is $\Omega_p/(2\pi) = 10\text{MHz}$, the intensity of the coupling field is $\Omega_c/2\pi = 10$ MHz, the attenuation of the intermediate state is $2\pi \times 5.2\text{MHz}$, the attenuation of the Rydberg state is $2\pi \times 1\text{MHz}$, and the detuning of the coupling field in the simulation result is $\Delta_c/(2\pi) = 0, \pm 25, \pm 50, \pm 75, \pm 100$ MHz , $\Delta_c/(2\pi) = 0, \pm 75$ MHz . The EIT signals at various detunings of the coupling light indicate that we may stimulate atoms in velocity groups corresponding to distinct detunings to the Rydberg state.

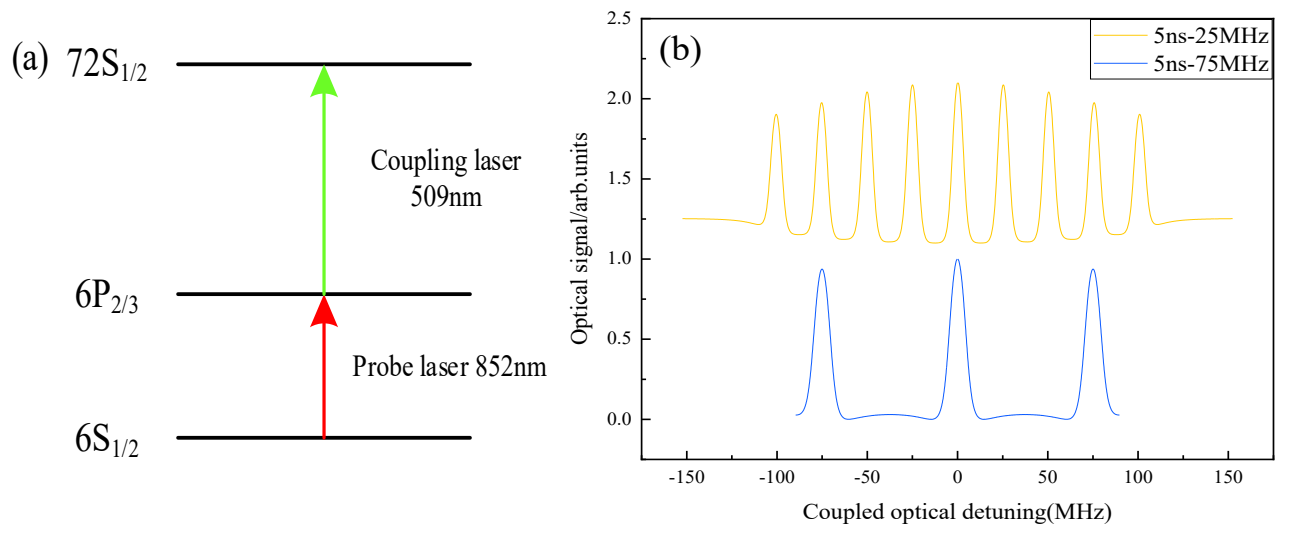

Fig. 1. (a) Rydberg EIT step-type energy level diagram of Cs atoms. (b) Theoretical simulation of EIT spectra with step-type multilevel structure where the coupled light is a pulsed laser, with coupled light detuning in the horizontal coordinates and normalized EIT transmission signal intensity in the vertical coordinates.

The response of the Rydberg atoms to the synthetic electric field after the signal field $E_{Sig}$ and the DC field to the electrodes are applied can be expressed as follows at this point:

$$E_{tot} = E_{DC} + E_{Sig}e^{i(\omega_{Sig}t + \Phi_{Sig})} \tag{6}$$

Among them, $E_{DC}$ and $E_{Sig}$ represent the amplitudes of the direct current (DC) field and the signal field respectively [19]. $\omega_{Sig}$ and $\Phi_{Sig}$ represent the angular frequency and the phase of the signal field respectively. This yields the squared amplitude of the synthesized electric field:

$$\begin{aligned} E_{tot} &= \left(E_{DC} + E_{Sig}e^{i(\omega_{Sig}t+\Phi_{Sig})}\right)\left(E_{DC} + E_{Sig}e^{-i(\omega_{Sig}t+\Phi_{Sig})}\right) \\ &= E_{DC}^2 + E_{Sig}^2 + 2E_{DC}E_{Sig}\cos\left(\omega_{Sig}t + \Phi_{Sig}\right) \end{aligned} \tag{7}$$

As $\Delta f_{Stark} = -\alpha E_{tot}^2/2$ is the Stark shift formula, where $\alpha$ denotes the polarizability. The Rydberg EIT signal oscillates at frequency $\omega_{Sig}/2\pi$, and its amplitude is linearly proportional to the signal field $E_{Sig}$. The oscillating signal is substantially enhanced by the DC field $E_{DC}$ during this procedure. We can measure the amplitude change of the oscillating signal by adjusting the intensity of the direct current (DC) field while maintaining the signal field unchanged in order to accomplish a highly sensitive measurement of the signal field. Ultimately, we determine the intensity of the DC field that optimizes the amplitude, thereby obtaining the most effective measurement effect [20].

Fig.2 illustrates the experimental apparatus. We choose lasers with wavelengths of 509 nm and 852 nm. We can effectively excite Cs atoms to the Rydberg state by meeting the condition of two-photon resonance with these two lasers. An external cavity semiconductor laser (ECDL) generates the 852 nm laser. The probe light is this 852 nm laser, and its frequency is locked at the transition of $6S_{1/2}$(F=4) → $6P_{3/2}$(F=5). The 852 nm laser is divided into two beams after passing through an optical isolator, a half-wave plate, and a polarization beam splitter (PBS). The frequency stabilization system, which is employed to stabilize the saturation absorption frequency of the 852 nm laser, receives one beam. The other beam, after passing through a half-wave plate and a PBS again, is split into two sub-beams. These two sub-beams respectively propagate collinearly in opposite directions with the two 509 nm laser beams. The EIT spectrum generated by one of the paths is used for the frequency stabilization of the 509 nm laser, and the other path is used for the electric field measurement experiment after exciting Cs atoms to the Rydberg state.

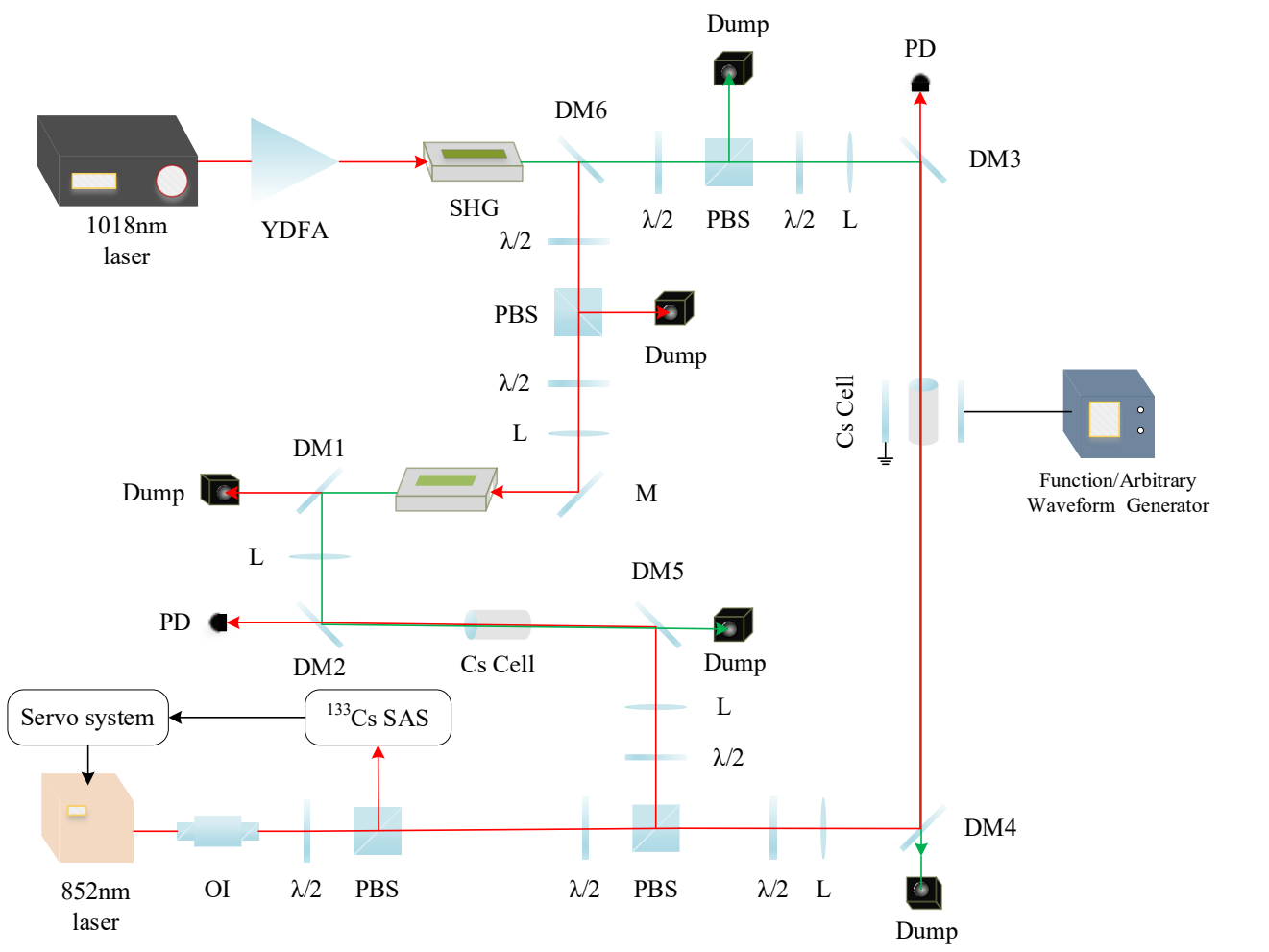

Fig. 2. Experimental setup. OI, optical isolator. YDFA, ytterbium-doped fiber amplifier. λ/2, half-wave plate. PBS, polarizing beam splitter cube. L, Lens. DM1, DM2, DM3, 509 nm high reflectivity (HR) and 852 nm high transmissivity (HT) dichroic mirror. DM4, DM5, 509 nm high transmissivity (HT) and 852 nm high reflectivity (HR) dichroic mirror. PD, photodiode. PPLN, periodically polarized lithium niobate crystals. M, 509 nm high reflectivity mirror. SAS, Cs atomic saturation absorption spectroscopy. Dump, optical dump

The pulsed light is subsequently amplified by a ytterbium-doped fiber amplifier (YDFA) and outputted to the system. We produce two beams of 509 nm pulsed laser light as coupling light by utilizing a self-built two-stage frequency doubling system. These beams are then directed into distinct Cs atomic vapor cells. Simultaneously, we apply a scanning voltage to the 1018 nm laser to regulate the frequency of the 509 nm coupling light within the $6P_{3/2}$ (F=5) $\rightarrow 72S_{1/2}$ transition. The cylindrical Cs atomic vapor cell utilized for the measurement has a diameter of 40 mm and a length of 70 mm. Within it, there are parallel electrode plates. The electrode is 60 mm in length, 15 mm in width, and 1.5 mm in height. The distance between the electrodes is approximately 10 mm. The electrode plates of the Cs atomic vapor cell are loaded with a voltage using a function waveform signal generator. The electric field signals were observed by means of photodetectors and spectrum analyzers.

The experiment involves a 509 nm pulsed laser with a pulse width of 5 ns and a repetition frequency of 75 MHz. Following the excitation of Cs atoms to the Rydberg state by the 509 nm and 852 nm lasers, we utilize the resultant EIT signal to stabilize the frequency of the 509 nm laser. Rydberg state Cs atoms can significantly enhance measurement sensitivity. Fig. 3(a) illustrates the generation of a Rydberg atom frequency comb under the influence of 509 nm pulsed light and an 852 nm laser. Different peaks correspond to atoms within distinct velocity groups, hence yielding Rydberg atoms at an increased density. Consequently, we proceed with the second task of measuring the electric field.

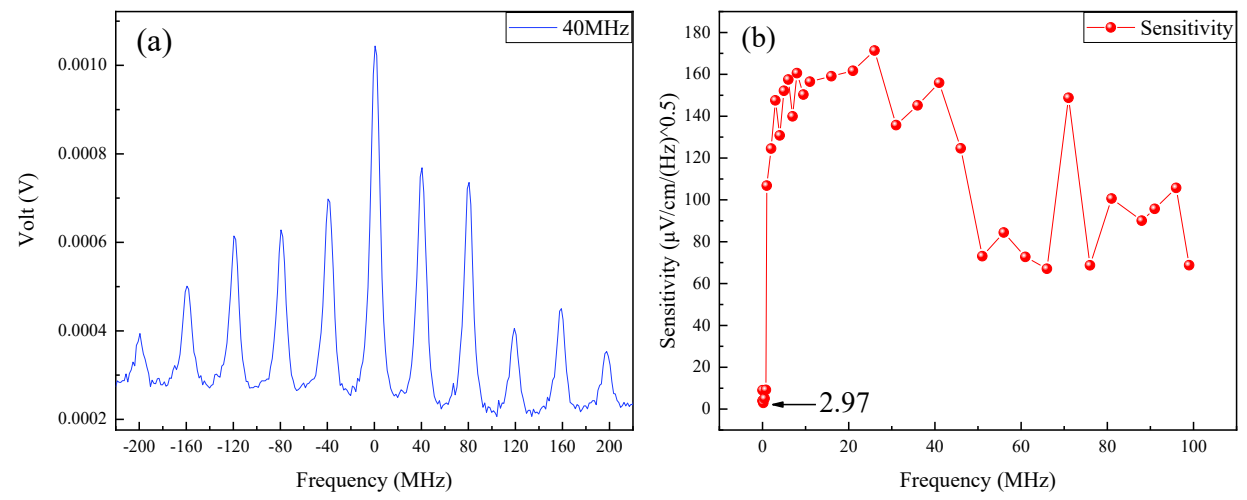

Fig. 3. (a) The EIT spectrum generated by the 509 nm laser with a pulse width of 4 ns and a repetition frequency of 40 MHz. (b) The sensitivity of electric field measurement in the frequency range of 20 kHz to 96 MHz.

In addition to the signal field, we introduce the Stark frequency shift by loading a direct current (DC) field onto the Cs atomic vapor cell and utilizing it as a local oscillator field. This results in a Rydberg energy level transition to a high sensitivity point, which in turn enhances the measurement sensitivity. In the experiment, it was discovered that a DC auxiliary field with an optimal intensity exists for electric fields of varying frequencies. This field can be used to accomplish the minimum necessary measurement sensitivity. The electric field measurement was completed in the 20 kHz to 96 MHz range and the corresponding measurement sensitivity was obtained, as illustrated in Fig. 3(b), following the addition of the DC auxiliary field with the optimal intensity. We can observe that the minimal measurement sensitivity is $2.9 \mu V/cm/Hz^{1/2}$.

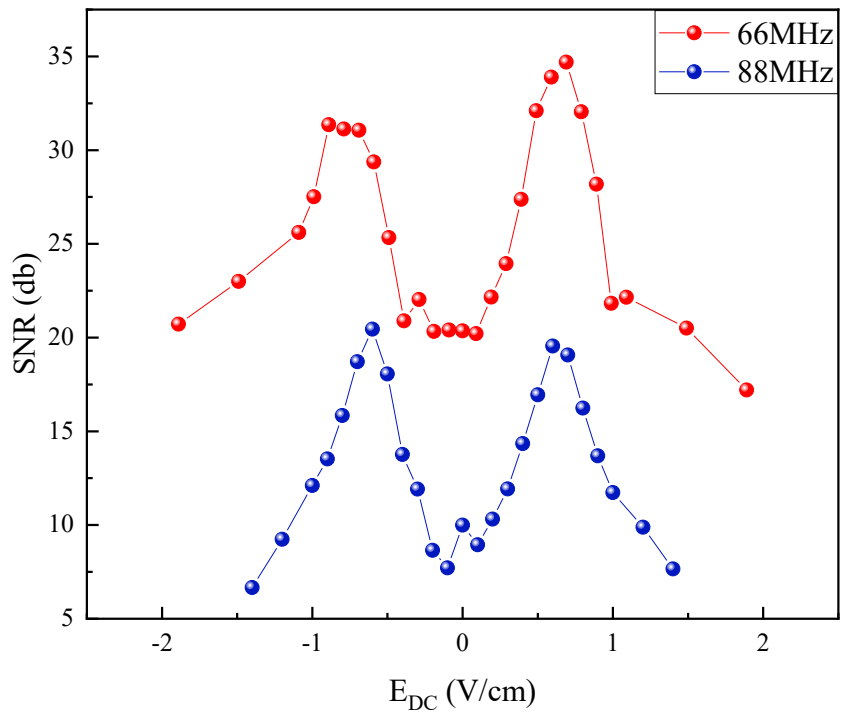

Fig. 4. Optimization of the auxiliary field for electric field frequencies of 66 MHz and 88 MHz.

The variation relationship between the signal field and the intensity of the auxiliary field at the broadcast communication frequencies of 66 MHz and 88 MHz has been measured, as illustrated in Fig. 4. As the intensity of the supplied auxiliary field varies, the enhancement effect on the signals also varies for signal fields of varying frequencies, as demonstrated in the experiment. There will always be an auxiliary field with an optimal intensity for each frequency, which can minimize the measurement sensitivity and obtain the highest SNR. The optimal intensity of the auxiliary field is 690 mV for the signal field with a frequency of 66 MHz. The optimal intensity of the auxiliary field is 600 mV for the signal field with a frequency of 88 MHz.

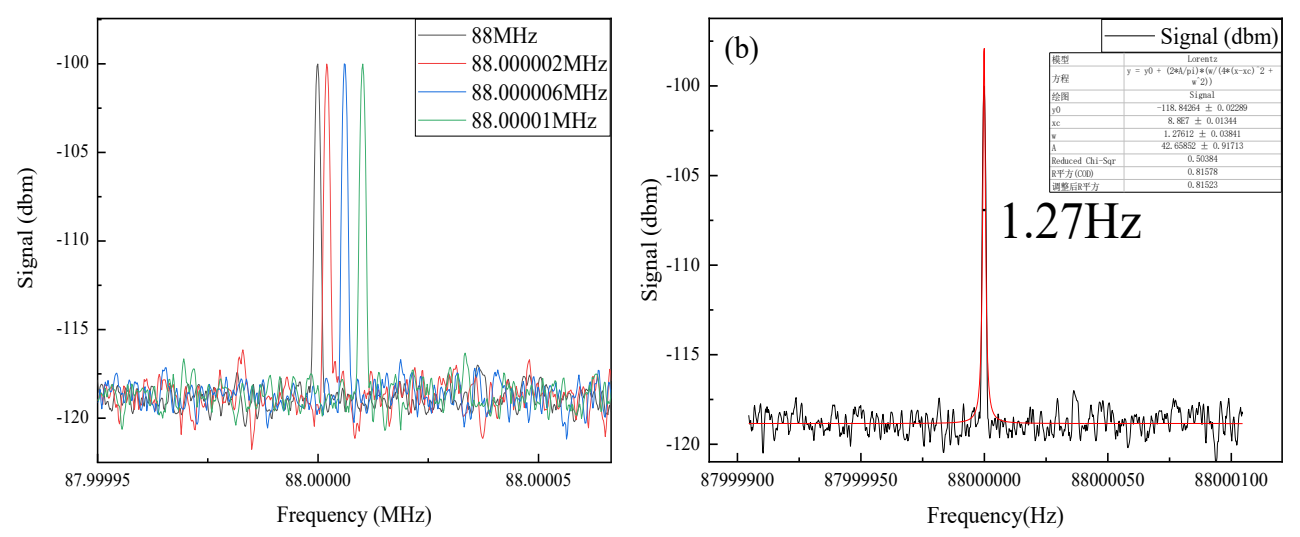

Fig. 5. (a) spectral resolution. (b) Lorentz peak fitting results.

High spectral resolution is critically important for accurate measurements using Rydberg atoms. It can markedly improve measurement sensitivity and accuracy, profoundly influencing the detection of weak signals. Fig. 5 illustrates that conducting a Lorentz peak fitting on the spectrum may significantly enhance data quality and analytical accuracy. It allows a more precise assessment of the position and strength of the exciton peak, thereby acquiring parameters such as peak position, intensity, and width. Fig. 5(b) illustrates that the spectral resolution of our established system may attain 1.27 Hz.

A parallel electrode plate-equipped Cs atomic vapor cell is employed to employ the Electromagnetically Induced Transparency (EIT) technique. Using a 509 nm pulsed laser and an 852 nm continuous-wave laser, we perform two-photon resonant transitions. This process effectively induces the $72S_{1/2}$ state Rydberg state in Cs atoms. Using a pulsed laser, we have successfully constructed an electric field sensor that is based on the Rydberg atom frequency comb. The precise measurement of electric fields across a broad frequency range from 20 kHz to 96 MHz is demonstrated by the application of an external DC field for optimization. The experimental results indicate that we can obtain a measurement sensitivity for the electric field as low as approximately $2.9\mu V/cm/Hz^{1/2}$ and a measurement sensitivity of microvolts per centimeter at the critical frequencies of 66 MHz and 88 MHz for broadcast communication.

The innovative aspect of this paper is the successful fabrication of a Rydberg atom frequency comb and the adoption of pulsed laser pumping technology to obtain Rydberg atoms. The pulsed laser is capable of exciting atoms within the entire velocity range, which increases the population of Rydberg atoms by two to three orders of magnitude in comparison to the traditional continuous laser excitation method. This substantially increases the density of sensor atoms, hence enhancing measurement sensitivity. Unlike the microwave frequency comb, our approach, utilizing the highly excited states of Rydberg atoms, may encompass the frequency spectrum from Radio Frequency (RF) to the terahertz band and beyond. Furthermore, the Rydberg atom frequency comb electric field sensor we created has exceptional measurement sensitivity at the broadcast communication frequency ranges of 66 MHz and 88 MHz. It integrates wide bandwidth and excellent sensitivity, indicating extensive application potential and considerable practical importance.

In conclusion, our research not only broadens the measuring frequency range of Rydberg atom electric field sensors but also attains exceptional sensitivity in electric field measurements. These findings offer novel technological methodologies and a theoretical basis for the utilization of quantum sensing technologies in precision measurement, broadcast communication, and other domains. Future research will enhance the utilization of Rydberg atom frequency combs in the terahertz frequency range and investigate measuring methodologies with increased sensitivity, with the objective of achieving the system's ultimate sensitivity.

**Funding.** This work is supported by the Natural Science Foundation of Chongqing of China (Grant No.CSTB2024NSCO-MSX0746), State Key and the Laboratory of Quantum Optics and Quantum Optics Devices (Grant No.KF202408), the National Natural Science Foundation of China (Grant No. 52175531).

**Disclosures**. The authors declare no conflicts of interest.

**Data Availability Statement (DAS).** A The data that support the findings of this study are available within the article.